\documentstyle[11pt,paspconf,epsf]{article}

\begin{document}

\title{Evolution in the Clustering of Galaxies for $Z < 1$}

\author{Robert J. Brunner}
\affil{Department of Astronomy, The California Institute of Technology,
Pasadena, CA 91125}

\author{Andrew J. Connolly}
\affil{Department of Physics and Astronomy, University of Pittsburgh,
Pittsburgh, PA, 15260}

\author{Alex S. Szalay}
\affil{Department of Physics and Astronomy, The Johns Hopkins University,
Baltimore, MD 21218}

\begin{abstract}

Measuring the evolution in the clustering of galaxies over a large
redshift range is a challenging problem. For a two-dimensional galaxy
catalog, however, we can measure the galaxy-galaxy angular correlation
function which provides information on the density distribution of
galaxies. By utilizing photometric redshifts, we can measure the
angular correlation function in redshift shells (Brunner 1997,
Connolly {\em et al.}  1998) which minimizes the galaxy projection
effect, and allows for a measurement of the evolution in the
correlation strength with redshift. In this proceedings, we present
some preliminary results which extend our previous work using more
accurate photometric redshifts, and also incorporate absolute
magnitudes, so that we can measure the evolution of clustering with
either redshift or intrinsic luminosity.

\end{abstract}


\section{Data}

The photometric data used in this analysis are located in the
intersection between the {\em HST} 5096 field and the CFRS 14 hour
field ({\em i.e.} the Groth Strip), covering approximately 0.054
Sq. Degree. All of the photometric data were obtained over several
observing runs using the Prime Focus CCD (PFCCD) camera on the Mayall
4 meter telescope at Kitt Peak National Observatory (KPNO).  All
observations were made through the broadband filters {$ U, B, R, \&\
I$}.

The photometric data were reduced in the standard fashion which is
detailed elsewhere (Brunner 1997). Source detection and photometry
were performed using SExtractor, which was chosen for its ability to
detect objects in one image and analyze the corresponding pixels in a
separate image. When applied uniformly to multi-band data, this
process generates a matched aperture dataset. Our detection image was
constructed from the {$ U, B, R, \&\ I$} images using a $\chi^2$
process (Szalay {\em et al.}  1998).

\section{Empirical Photometric Redshifts}

The photometric redshifts used in this analysis were derived
empirically using a piecewise approximation approach (Brunner {\em et
al.} 1999). Briefly, this approach defines five different redshift
intervals which track the movement of the 4000 \AA\ break through our
filter system with increasing redshift (for $z < 1.2$). 190
calibrating redshifts were used to derive a global third order
polynomial in {$ U, B, R, \&\ I$} which provided an initial redshift
estimate, from which the appropriate local redshift estimate was
selected. The range of calibrating redshifts for each polynomial fit
was extended by approximately 0.05 in order to diminish end-aliasing
effects. This algorithm is designed to generate an optimal redshift
for objects by using the more accurate local relations (Brunner {\em
et al.} 1997).  For each derived polynomial fit, the degrees of
freedom remained a substantial fraction of the original data (a second
order fit in four variables requires 15 parameters while a third order
fit in four variables requires 35 parameters).

\begin{figure}[bth]
\plottwo{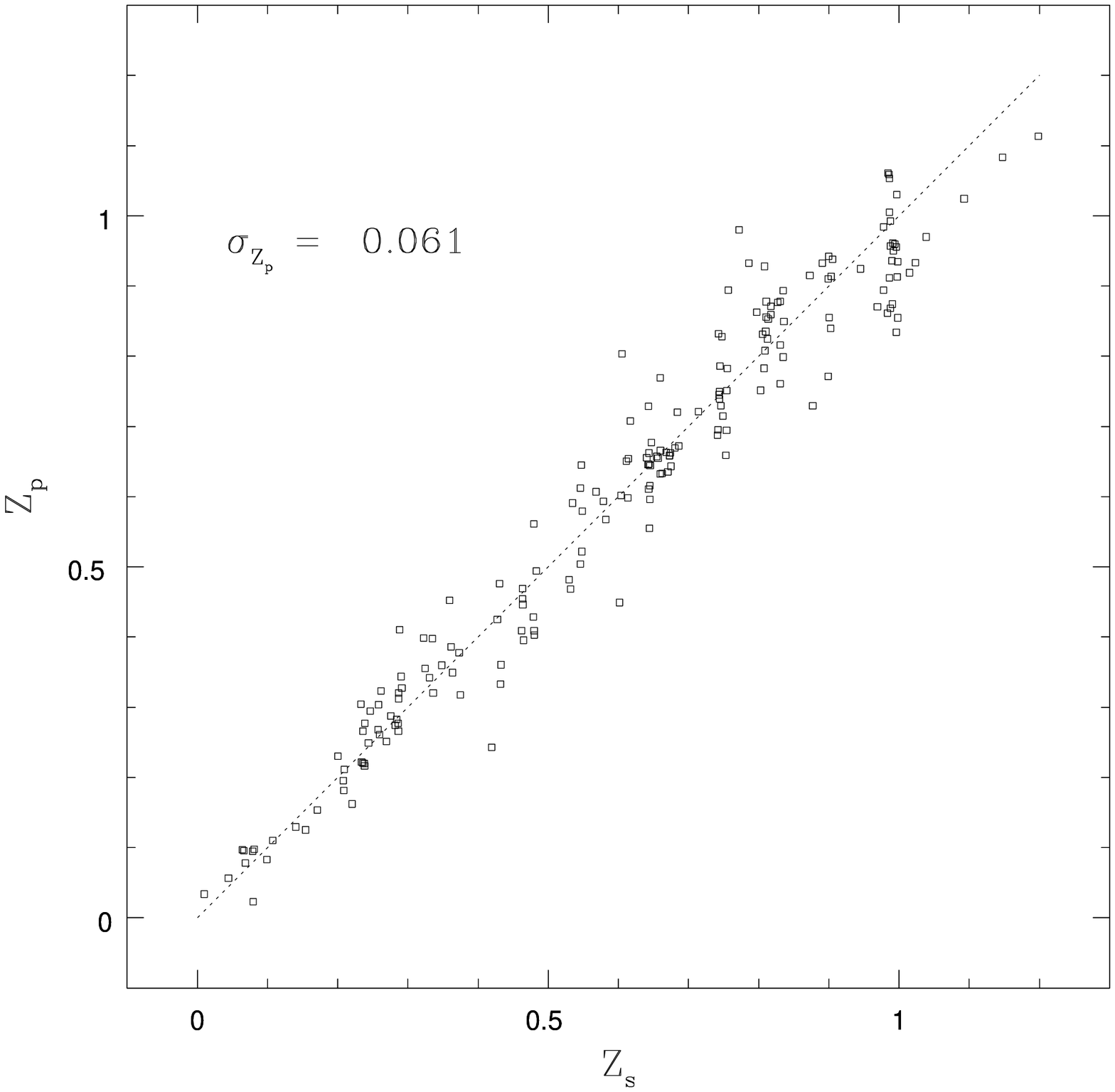}{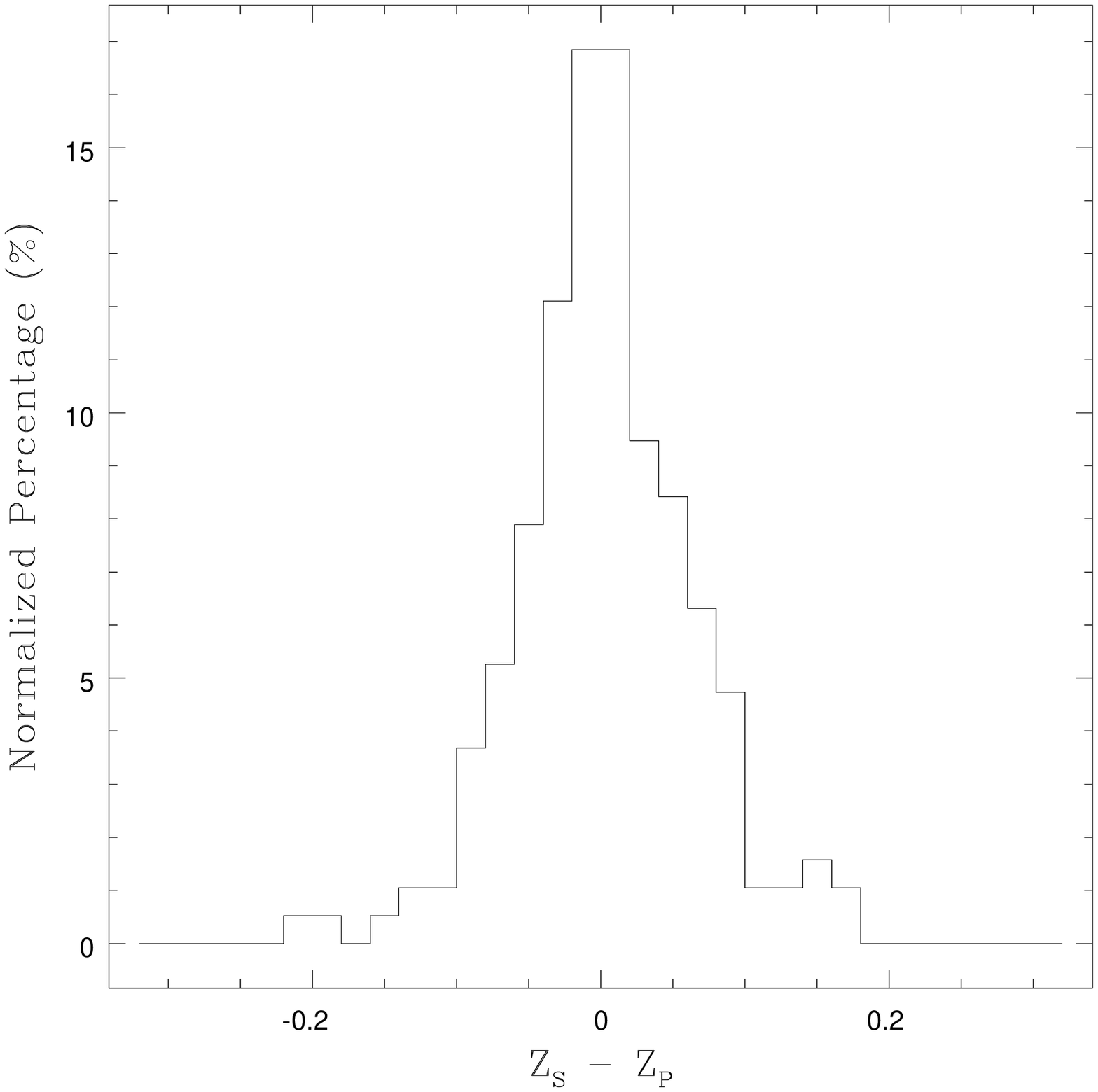}
\caption{The left hand figure provides comparison between the 
calibrating spectroscopic redshifts and the estimated photometric
redshifts used in this analysis. The intrinsic dispersion in the
relationship is $\sigma = 0.061$. The right hand figure provides a
histogram view of the residual differences between the spectroscopic
and photometric redshifts.
\label{ZP}}
\end{figure}

A subtle, and often overlooked, effect in any photometric redshift
analysis is the requirement for accurate multi-band
photometry. Ideally we could restrict our photometric-redshift catalog
to only those objects which have measured magnitude errors below some
set limits ({\em e.g.,} $10\%$ photometry). This type of a
restriction, however, introduces two complications: a bias towards
blue spectral types, and a subsequently complicated selection effect.

In an attempt to overcome these biases, we restrict the full sample to
those objects which have both $I_{AB} < 24.0$ and measured magnitude
errors $< 0.25$ in {$ U, B, \&\ R$}. This minimizes any selection bias
to only faint early-type galaxies.  The remaining filter combinations
contribute to the noise in our analysis ({\em i.e.} when we consider
our final catalog complete to $I \approx 24.0$), and amount to only a
few percent when combined. After removing stars and sources with bad
detection flags, our final photometric-redshift catalog contains 3052
sources.

\section{Ensemble Approach}

A formal, analytic technique is not always available to utilize
photometric redshifts and their associated errors when measuring
cosmologically interesting quantities. As a result, we have developed
an alternative technique, the galaxy ensemble approach. Essentially,
we treat the problem in the context of statistical mechanics, where
each galaxy is localized in redshift space by a Gaussian probability
distribution function. To calculate a physically meaningful quantity,
we create multiple realizations (or ensembles) of the galaxy redshift
distribution, and calculate the appropriate quantity for all of the
different ensembles. We then average the different measurements to
produce the desired value, simultaneously producing a realistic error
estimate. This can easily be done for the redshift distribution of the
galaxy sample, where an analytic approach is also available for
comparison ({\em cf.} Brunner {\em et al.} 1999).

This approach was used to generate absolute magnitude distributions in
both the U and B bands for this data. First, a set of 100 redshifts
were estimated for each source by drawing redshifts randomly from a
Gaussian probability distribution function (PDF) with mean and sigma
given by the calculated photometric redshift and photometric redshift
error. For each redshift in the ensemble, a k-correction was
calculated using the assigned spectral type. Apparent magnitudes were
estimated for every source in the ensemble by drawing them from a
Gaussian PDF with mean and sigma given by the measured magnitude and
magnitude error. Using the k-corrections, apparent magnitudes, and an
assumed cosmology, 100 different absolute magnitude distributions in
both B and U were generated.

\section{Angular Correlation Function}

Before computing the angular correlation function, we determined the
regions within our image in which the detection efficiency might be
reduced. The primary areas where this occurs are around bright stars,
in charge transfer trails, and near the edge of the frame due to edge
effects or focus degradations. We, therefore, defined bounding boxes,
for each of the four stacked images, which contained all of the
observable flux for the saturated stars within the image. These four
mask files were concatenated to produce a total mask file which was
used for the calculation of the angular correlation function in
different redshift intervals.

We used the optimal estimator $(DD - 2DR + RR)/RR$ (Landy \& Szalay
1993), where $D$ stands for data and $R$ stands for random, to
determine the angular correlation function. This required counting the
number of observed pairs (that were not within the masked areas),
which was done in 10 bins of constant width $\Delta\lg(\theta) =
0.25$, centered at $\theta = 4.3\arcsec$, to $\theta = 759.6\arcsec$.
1000 objects were then randomly placed in the non masked areas within
the image, and the data-random and random-random correlation functions
were calculated for the same angular bins used for the data-data
auto-correlation function.

This estimator uses the calculated number density of galaxies within
the CCD frame to estimate the true mean density of galaxies. The small
angular area of our images introduces a bias in the estimate, commonly
referred to as the ``Integral Constraint'' (Peebles 1980). We
estimated a correction for this bias following the prescription of
Landy and Szalay (1993), which is subtracted from the $(DD - 2DR + RR)/RR$
value.

The error in the estimation of the angular correlation function is
assumed Poisson in nature, and is calculated as the square root of the number
of random-random pairs in each angular bin.

\section{Evolution in the Clustering of Galaxies}

The multivariate angular correlation function $w(\theta, z_{P})$ was
determined for nine different redshifts by binning the data in
redshift bins of width $\Delta z_{P} = 0.2$ centered at $z_{P} = 0.2$
to $z_{P} = 1.0$. The nine different functions are calculated for the
258, 330, 327, 420, 640, 715, 713, 732, and 514 objects in the
different respective redshift bins. For each redshift region, a least
squares fit was performed assuming the relation $w(\theta) = A_{w}
\theta^{-0.8}$. The amplitude was then measured by minimizing the
absolute deviation with respect to $A_{w}$, which in this case reduces
to finding the median of the measured correlation function amplitudes.

Previously, two methods have been used to quantify the evolution in
the clustering of galaxies. The first technique is to invert the
angular correlation function ($w(\theta)$) using the Limber equation
(Peebles 1980) and an observed or model redshift distribution to
estimate the expected change in the amplitude of the angular
correlation function for different magnitude intervals and/or
cosmologies.  The other approach is to compute the spatial correlation
function ($\xi(r)$) for different epochs directly using spectroscopic
redshifts. These two techniques, however, suffer from different
limitations that have restricted their utility.

Studies which utilize different magnitude intervals are limited by the
redshift smoothing effects of an apparent magnitude limited sample,
much the same as the number magnitude distributions. The spectroscopic
approach is hindered either by the size of the available samples,
especially when the data is binned into distinct redshift intervals,
or by the depth of the survey. These studies are also affected (up to
a factor of two) by the redshift space distortions due to density
inhomogeneities.

\begin{figure}[bth]
\plottwo{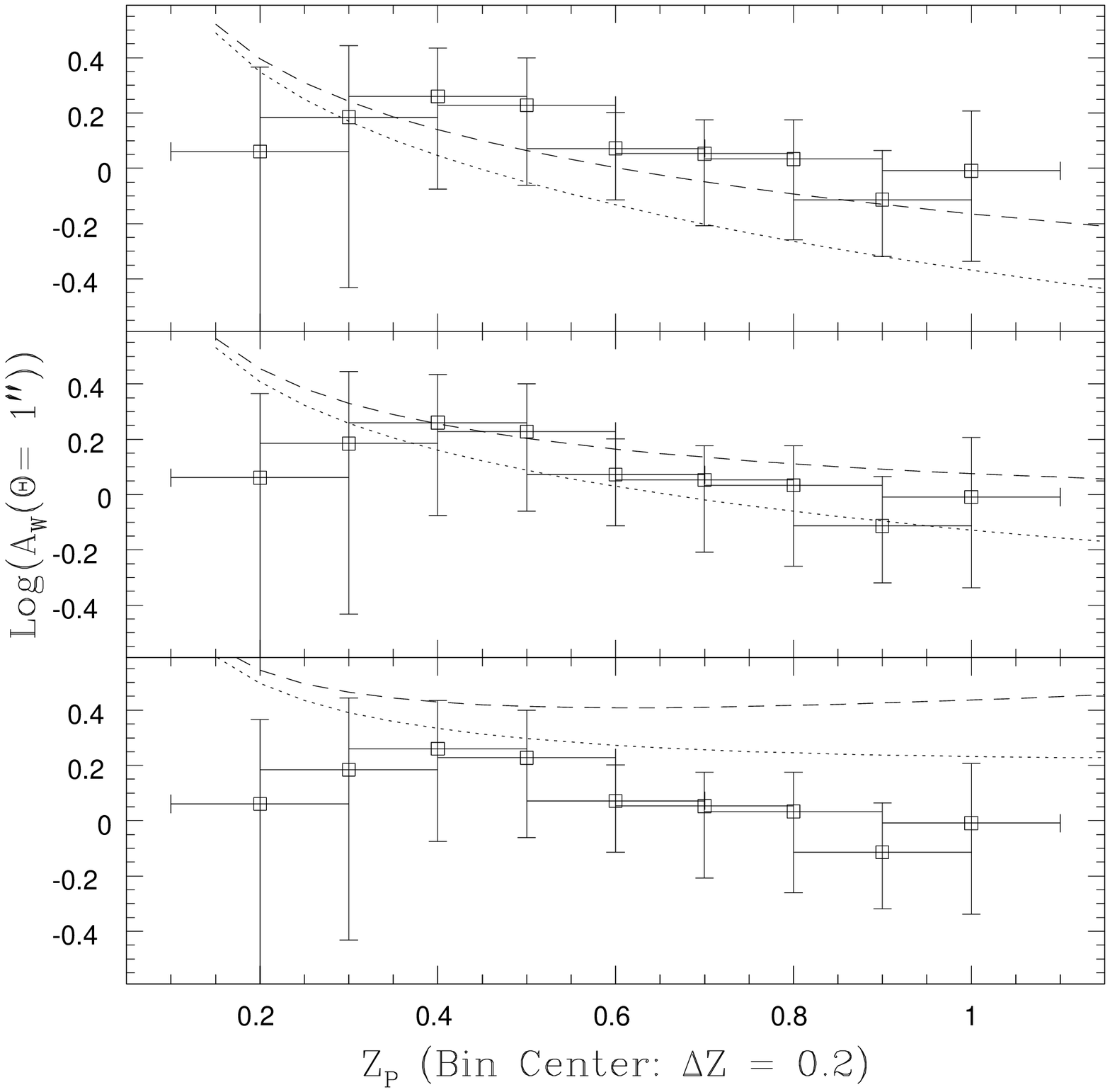}{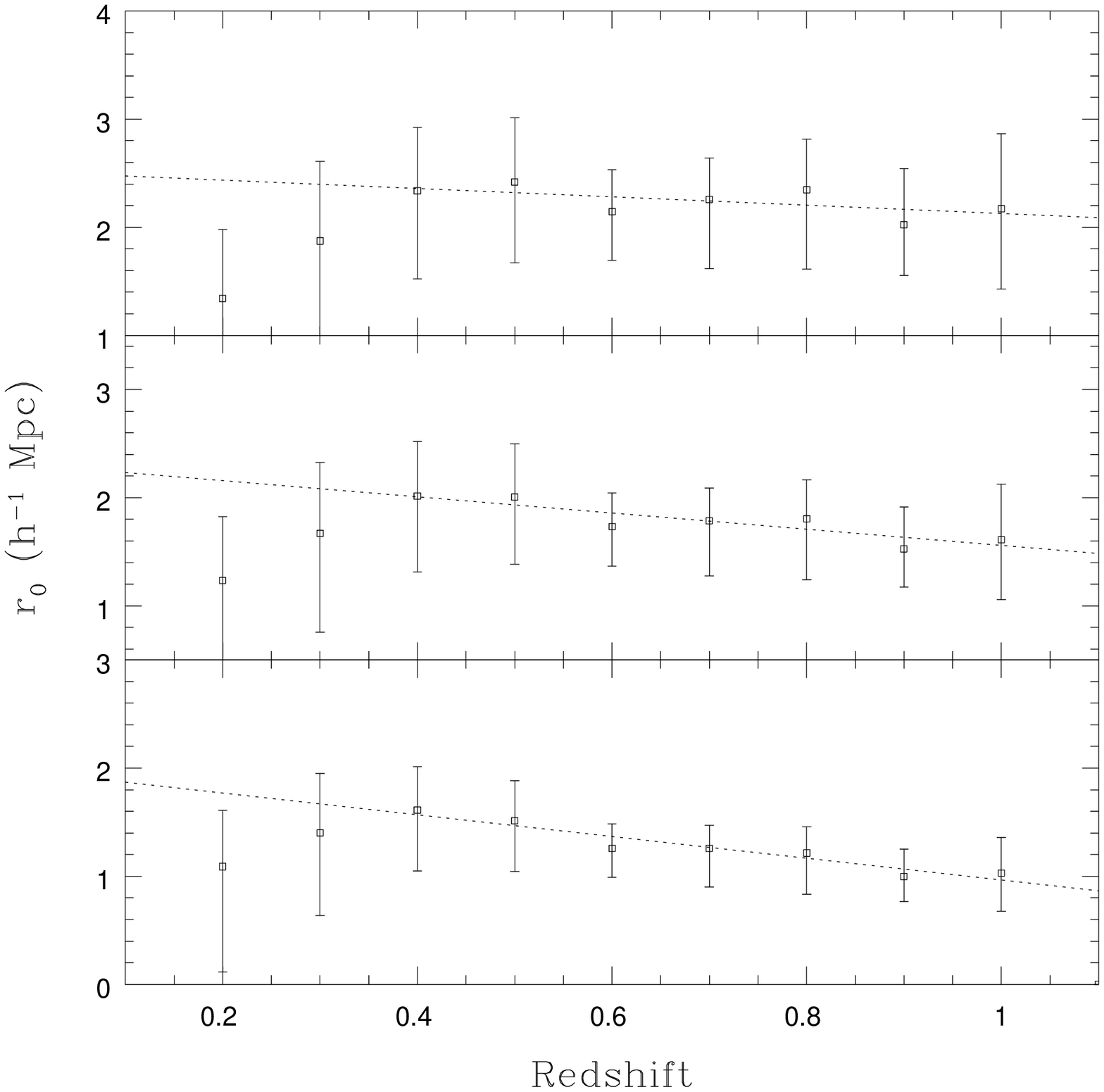}
\caption{The left hand figure displays the evolution in the amplitude 
of the angular correlation function with redshift. The three lines are
predictions for $\Omega = 0.1$ (dotted line) and $\Omega = 1.0$
(dashed line) using Limber's equation. The right hand figure shows the
evolution in the correlation length ($r_0$) with redshift assuming
$\Omega = 1.0$ using Limber's equation and the redshift distribution
measured from the galaxy redshift ensemble. In each figure, the top
panel assumes the evolution parameter derived from linear theory
($\epsilon\ = 0.8$), the middle panel assumes fixed clustering in
proper coordinates ($\epsilon\ = 0.0$), and the bottom panel assumes
fixed clustering in co-moving coordinates ($\epsilon\ = -1.2$).
\label{R0vZ}}
\end{figure}

We adopt the novel approach of computing the amplitude of the
multivariate angular correlation function for the 3052 objects in the
photometric redshift--template SED catalog in redshift bins of width
$\Delta z = 0.2$. To compare these observations to semi-analytic
theory, we assume a power law model for the spatial clustering
(Peebles 1980),
\[
\xi(r,z) = \left( \frac{r}{r_{0}} \right )^{-\gamma} (1 + z)^{-(3 + \epsilon)}
\]
where $\gamma \approx 1.8$ and $r_{0} \approx 3.0 \ h^{-1}$ Mpc in
co-moving coordinates as measured locally. The three canonical values
of the evolution (assuming $\gamma = 1.8$) are $\epsilon = 0.8$ as
predicted by linear theory, $\epsilon = 0.0$ for clustering fixed in
proper coordinates, and $\epsilon = -1.2$ for clustering fixed in
co-moving coordinates. The change in $A_{w}$ with redshift is
displayed in Figure~\ref{R0vZ} along with the three different
evolution models for $\Omega = 1.0$ and $\Omega = 0.1$. The error in
$A_{w}$ was calculated by estimating $A_{w}$ in each redshift interval
for the $1
\sigma$ upper and lower values.

Of the three different scenarios, the predictions for clustering fixed
in co-moving coordinates ($\epsilon = -1.2$) are the least consistent
with our data, independent of the value of $\Omega$. The predictions
of linear theory ($\epsilon = 0.8$) are mildly consistent for high
values of $\Omega$, while the best agreement is for clustering fixed
in proper coordinates ($\epsilon = 0.0$), independent of the value of
$\Omega$. Our technique is less sensitive to redshift distortions than
the spatial correlation approach due to the width of our redshift
bins. Furthermore, our technique does not require model predictions
for the redshift distribution of galaxies as does the apparent
magnitude interval approach.

\begin{figure}[t]
\plottwo{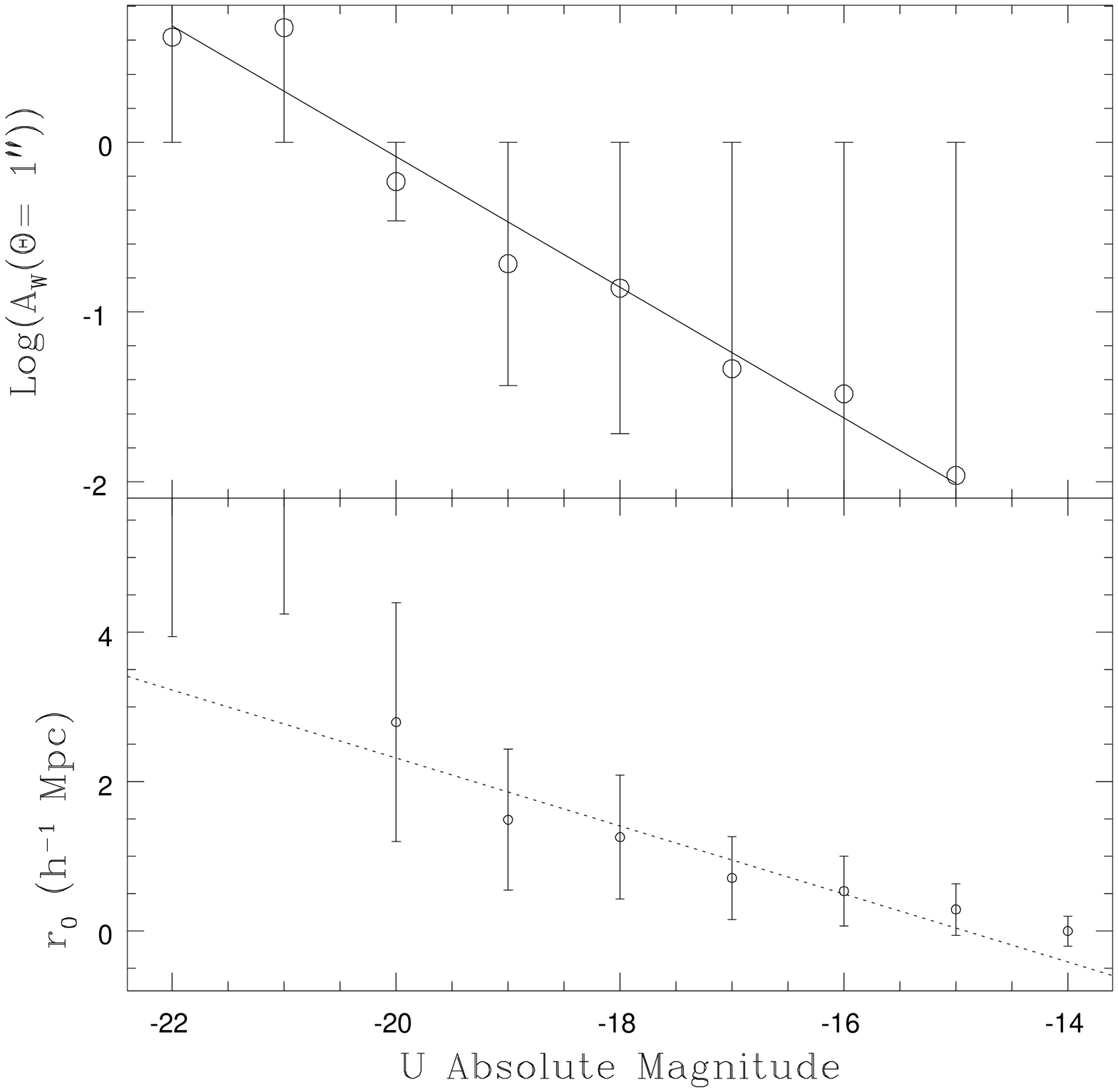}{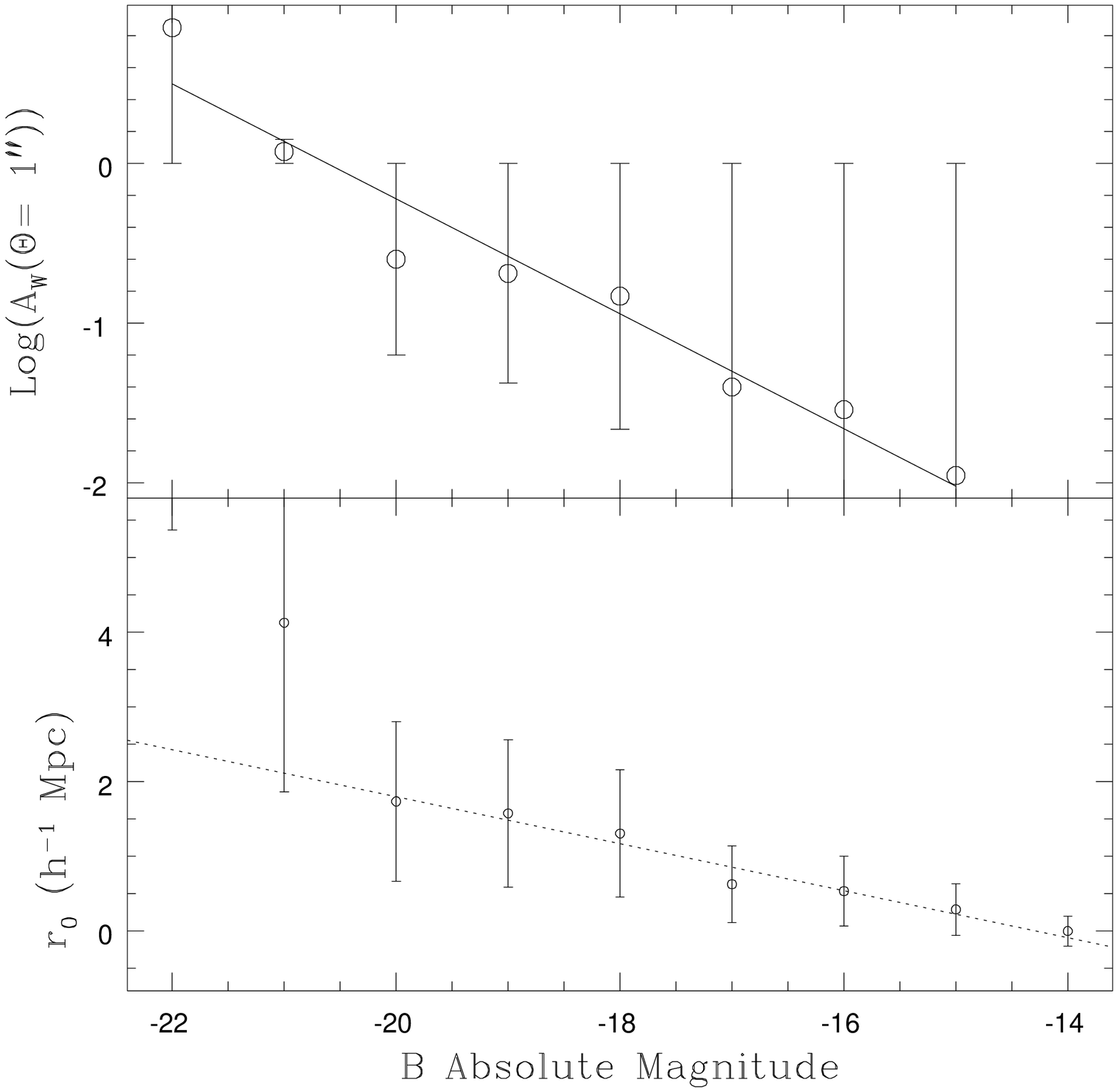}
\caption{These two panels show the evolution of the amplitude of 
the angular correlation function with absolute magnitude (top) and the
evolution of the correlation length with absolute magnitude (bottom)
for the U (right) and B (left) bands. No segregation by redshift was
used in this analysis. The absolute magnitudes are generated using the
100 different realizations with the ensemble approach. The amplitude
of the correlation function was converted to $R_0$ using Limber's
equation with $\Omega = 1$ and $\epsilon = 0.0$ and integrating over
entire redshift range.
\label{zAw}}
\end{figure}

The novel measurement of the evolution in the clustering of galaxies
with absolute magnitude we have presented in Figure~\ref{zAw} is
consistent with both the general expectation of most structure
formation theories and similar low redshift observations. Further
improvements in this technique are forthcoming (Brunner {\em et al.}
1999b), and will include measurements of the evolution of the
correlation length with both redshift and absolute magnitude, as well
as a conversion from the amplitude of the correlation function ($A_w$)
to the correlation length ($r_0$) in a model ({\em i.e.} $\epsilon$)
free technique.

\acknowledgments

We wish to thank George Djorgovski, Mark SubbaRao, Gyula Szokoly, and
Lori Lubin for useful Discussion. AJC acknowledges partial support
from HST (GO-07817-02-96A) and LTSA (NRA-98-03-LTSA-039), AS from NASA
LTSA (NAG53503), HST Grant (GO-07817-04-96A).

\end{document}